\documentclass[conference]{IEEEtran}
\IEEEoverridecommandlockouts
\usepackage{cite}
\usepackage{amsmath,amssymb,amsfonts}
\usepackage{algorithmic}
\usepackage{graphicx}
\usepackage{textcomp}
\usepackage{xcolor}
\usepackage{setspace}

\usepackage{algorithmic}
\usepackage{booktabs}  
\usepackage{threeparttable} 
\usepackage{multirow}
\usepackage{booktabs}
\usepackage[misc]{ifsym}

\def\BibTeX{{\rm B\kern-.05em{\sc i\kern-.025em b}\kern-.08em
		T\kern-.1667em\lower.7ex\hbox{E}\kern-.125emX}}

\begin{document}

\title{DT-SV: A Transformer-based Time-domain Approach for Speaker Verification 
	\thanks{\dag \quad Corresponding Author: Jianzong Wang, jzwang@188.com}}
\author{\IEEEauthorblockN{Nan Zhang, Jianzong Wang\textsuperscript{\dag}, Zhenhou Hong, Chendong Zhao, Xiaoyang Qu, Jing Xiao}
	\IEEEauthorblockA{\textit{Ping An Technology (Shenzhen) Co., Ltd., Shenzhen, China} \\
		{Emails: nzhang889@gmail.com, jzwang@188.com,} \\  
		{\{zhenhouhong, cd896614, quxiaoy\}@gmail.com, xiaojing661@pingan.com.cn}}
}

\maketitle

\begin{abstract}
Speaker verification (SV) aims to determine whether the speaker’s identity of a test utterance is the same as the reference speech. In the past few years, extracting speaker embeddings using deep neural networks for SV systems has gone mainstream. Recently, different attention mechanisms and Transformer networks have been explored widely in SV fields. However,  utilizing the original Transformer in SV directly may have frame-level information waste on output features, which could lead to restrictions on capacity and discrimination of speaker embeddings. Therefore, we propose an approach to derive utterance-level speaker embeddings via a  Transformer architecture that uses a novel loss function named \textit{diffluence loss} to integrate the feature information of different Transformer layers. 
Therein, the \textit{diffluence loss} aims to aggregate frame-level features into an utterance-level representation,  and it could be integrated into the Transformer expediently. Besides, we also introduce a learnable mel-fbank energy feature extractor named \textit{time-domain feature extractor} that computes the mel-fbank features more precisely and efficiently than the standard mel-fbank extractor. Combining \textit{Diffluence loss} and \textit{Time-domain feature extractor}, we propose a novel Transformer-based time-domain SV model (DT-SV) with faster training speed and higher accuracy. Experiments indicate that our proposed model can achieve better performance in comparison with other models.

\end{abstract}


\section{Introduction}
\label{sec:intro}
Speaker verification (SV) is a binary classification task that answers the question whether an unknown utterance belongs to its claimed identity. Usually, it can be divided into two categories: text-dependent speaker verification (TD-SV) and text-independent speaker verification (TI-SV)\cite{qu2020evolutionary}. Therein, TI-SV has no constraint on text content and speakers can say anything to the verification system, which brings great convenience to the users, hence we focus on TI-SV in this work.



Generally speaking,  there are two kinds of models in the research of SV: the statistical model and the neural network model\cite{mingote2021memory}. As an efficient statistical model, i-vector\cite{dehak2010front} achieves great success in TI-SV task. It compresses both speaker and channel information into a fixed-dimensional space called total variability subspace. Recently, with the increasing of the scale of labeled data, more and more researchers start to pay attention to the neural network model. Under the supervised learning framework, the model could automatically learn the speaker representation through the data-driven training method.

Deep neural networks have been shown to be useful for extracting speaker-discriminative feature vectors independently from the i-vector framework~\cite{dehak2010front}. 
With the help of an amount of training data, such approaches have obtained much better results, particularly under the condition of short-duration utterances. 
In 2014, Ehsan et al.\cite{variani2014deep} introduced the neural network into speaker verification. At the training step, four fully connected layers are utilized for speaker classification. Meanwhile, the speaker embedding (‘d-vector’) is calculated from averaging the last hidden layer’s output over frames in verification step. Using this pipeline, more well-designed neural networks such as convolutional neural network(CNNs)\cite{balian2021small,han2021time} and recurrent neural networks(RNNs)\cite{tak2021end} have been proposed for SV task. However, it still needs more powerful deep neural networks to better extract the speaker feature.



The attention mechanism\cite{bahdanau2015neural, zhang2021cacnet} is a powerful method which offers a way to obtain an even more discriminative utterance-level feature by explicitly selecting frame-level representations that better represent speaker characteristics. 
Nowadays, Transformer with self-attention mechanism have become an effective model in a variety of application fields\cite{india2021double, wu2021adversarial, hong2021federated},  which mainly focus on the processing of deep neural networks on certain areas of feature maps or certain temporal slots, including the scenario of SV\cite{mingote2021memory}. The success achieved by this approach in SV with self-attention mechanism allows models to learn the frame-level features, which are more precise to represent the speaker characteristics. Yet, there are still two issues. First, since it can not extract speaker-discriminative features from the raw speech input, it relies on the mel-fbank or MFCC features. Secondly, the original Transformer structure may have frame-level information waste on output features, which could lead to restrictions on capacity and discrimination of speaker embeddings.

Based on above issues, we propose a Transformer-based Time-domain model for SV named DT-SV with two major improvements in comparison to the original Transformer in SV, a novel loss function named \textit{diffluence loss} and a learnable mel-fbank feature extractor named \textit{time-domain feature extractor}.  Features extracted by this extractor will be more appropriate to the neural network since the \textit{time-domain feature extractor} is learned from the raw data distribution. Also, \textit{diffluence loss} contributes to better obtain the utterance-level embedding, which summarizes the information from other embeddings via a self-attention mechanism. Specifically, this loss indicates the distance among features of the first frame and other frames on each layer, thus enhancing the speaker-related information in the utterance-level embedding while weakening the speaker-related information among other frame embeddings. In particular, we propose two architectures with different numbers of layers, which are DT-SV-light and DT-SV. Both of them are well-performed in our experiments while the DT-SV-light could achieve competitive performance compared to other light models with 10 times smaller GFLOPs.

Our main contributions are as follows:
\begin{itemize}
\item An utterance-level speaker representation based on Transformer is presented, via proposing a novel \textit{diffluence loss} to aggregate frame-level features on each layer to an utterance-level speaker representation while weaken the speaker-related information in the frame-level embeddings.
\item A learnable mel-fbank features extractor named \textit{time-domain feature extractor} is introduced, which could extract features from speech signal like mel-fbank more precisely and efficiently than the standard mel-fbank extractor. 
\item A Transformer-based time-domain speaker verification model is designed to take advantage of the above two modules, achieving faster training speed and higher accuracy. 

\end{itemize}

\section{Related Work}
\label{sec:relatedwork}
\subsection{Speaker Embedding Extractor}
During the last few years, how to obtain a single fixed dimension vector to represent an utterance has drawn most researchers' attention in this field.
The success of i-vector\cite{dehak2010front} extracted from GMM-UBM models\cite{reynolds2000speaker}, one of the earliest speaker embeddings, stimulated the interest among researchers to search for better speaker embeddings. 
In the meantime, with the increase of available data for training, speaker embeddings based on deep learning methods have gained popularity.
The d-vector\cite{variani2014deep} is one of the earliest DNN-based embeddings, the core idea of which is to assign the ground-truth speaker identity of an utterance as the labels of frames belonging to this utterance in the training phase, which transfers the model training to a classification problem.
Soon after, the x-vectors\cite{snyder2017deep, snyder2018x} extracted from the Time Delay Neural Networks(TDNNs) \cite{waibel1989phoneme} based system have consistently outperformed i-vector based systems and therefore been widely used in state-of-the-art systems. Considering spectrogram representations of speech as images, speaker embeddings extracted from pretrained residual networks (ResNets) \cite{nagrani2020voxceleb,wang2021hierarchically} with a considerable number of images, have performed even better than x-vectors recently. 
In order to represent variable utterance lengths, different pooling methods \cite{chung2018voxceleb2,snyder2018x,okabe2018attentive} have been proposed to aggregate the information across the utterance into a single vector. The s-vectors \cite{nj2021s} extracted from speaker classification architectures based on Transformer’s encoder, attend to all features that capture speaker characteristics over the entire utterance, achieving greater performance than x-vectors.

\subsection{Transformer in Speech}
The attention-based model has nowadays shown its powerful ability on representation learning in speech processing fields. There have been a few studies to explore incorporating attention at the model level and prove such attention mechanism is effective in aggregating frame-level features.
Zhang et al.\cite{ zhang2020speaker}  proposed the time-frequency attention module for speaker verification to help the system to focus on important regions of short utterances along the time and the frequency domain, extracting more effective speaker discriminative information. In \cite{ wang2021hierarchically}, an end-to-end speaker verification system employing time-frequency and channel attention hierarchically in ResNet framework is presented for speaker verification as well. 

Lately, the architecture named Transformer with stacked self-attention at the encoder and a multi-head topology have been explored extensively for SV. Inspired by the next sentence prediction task of BERT\cite{devlin2018bert}, an architecture called TESA \cite{nj2021s} based on Transformer’s encoder is proposed as a replacement for conventional PLDA-based speaker verification to capture the speaker characteristics better, outperforming the PLDA\cite{ioffe2006probabilistic} on the same dataset.
Besides, Zhu et al.\cite{zhu2021serialized} proposed a method to create fixed-dimensional representations for SV with a serialized multi-layer multi-head attention mechanism containing a stack of self-attention modules, which is designed to aggregate and propagate attentive statistics from one layer to the next in a serialized manner.  Different from other studies which re-designs the inner structure of the attention module, we strictly follows original Transformer, providing the simple but effective modifications.

\begin{figure*}
	\centering
	\includegraphics[scale=0.45]{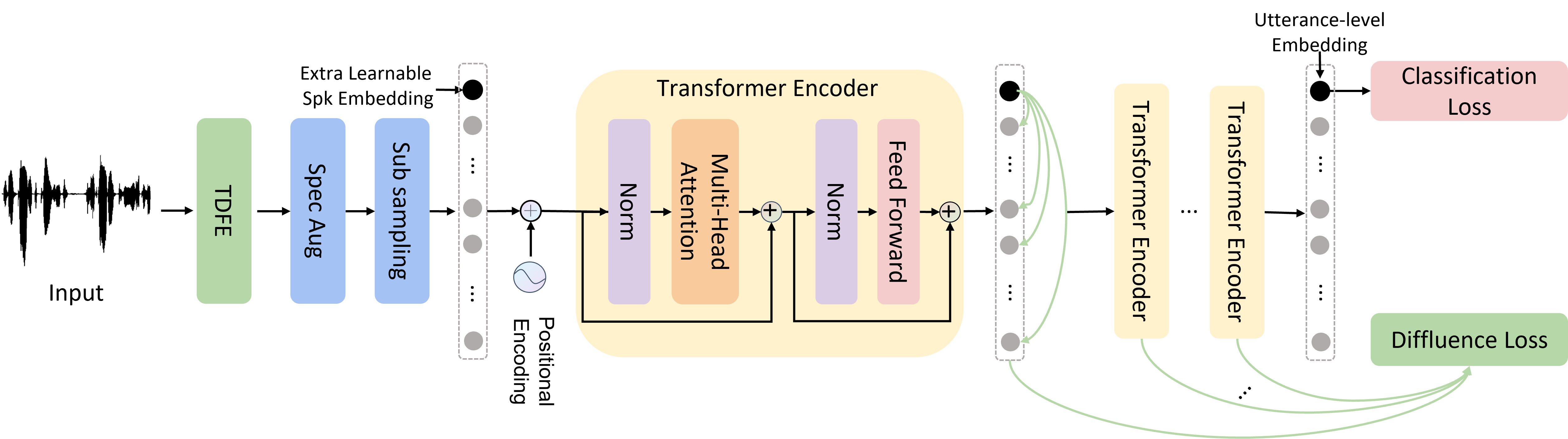}
	\caption{ A diagram of DT-SV, which is based on Transfomer with two major improvements in green part. Therein, the Transformer block is the same as the usual Transformer layer and TDFE represents Time-domain Feature Extractor.}
	\label{fig:transformer-spkformer}
\end{figure*}

\section{Proposed Method}
\label{sec:model}
\subsection{Model Architecture}
A common practice in all deep learning methods addressing the SV task is to perform speaker classification first. Then, utterance-specific fixed-dimension embeddings are obtained from the speaker classification network by different pooling methods. These embeddings are immediately fed to a SV system to validate the identity claimed by the speaker. 

In the framework based on the original transformer structure in SV, the input is mel-fbank features while the output is the speaker classification loss. Similar to the Transformer applied to natural language processing like BERT\cite{devlin2018bert} and computer vision like ViT\cite{dosovitskiy2020image}, an extra learnable embedding is added to the sequence as the first frame embedding in our proposed model. The final hidden state corresponding to this frame serves as the aggregate utterance-specific embedding for speaker classification, ignoring the information in other frames.
However, we noticed that the speaker-related information not only exists in the first frame embedding but also is contained in embeddings of other ignored frames via our experiments.  For the better capacity and discrimination of speaker embeddings, the ignored information should be further aggregated to form an utterance-level representation. To address this issue, we proposed a novel SV model based on the original Transformer with some minor modifications on loss function.

As illustrated in Fig.\ref{fig:transformer-spkformer}, the major differences between our proposed DT-SV and the original Transformer-based model are \textit{time-domain feature extractor}(TDFE) and \textit{diffluence loss}. Therein, TDFE is a front module to extract features from the raw waveform input directly, and \textit{diffluence loss} is the sum of distances between the utterence-level embedding and each frame-level embedding in every layer.
Particularly, such loss utilizes all features over the entire utterance, which is more suitable in capturing speaker characteristics on an utterance. Meanwhile, the join of DTFE makes DT-SV completely an end-to-end SV model directly from a raw input to classification results, which accelerates forward propagation speed and model training speed since all operations could be implemented on GPU because of the raw time domain input.

\subsection{Time-domain Feature Extractor}

\begin{figure}[htbp]
	\centering
	\includegraphics[width=0.9\linewidth]{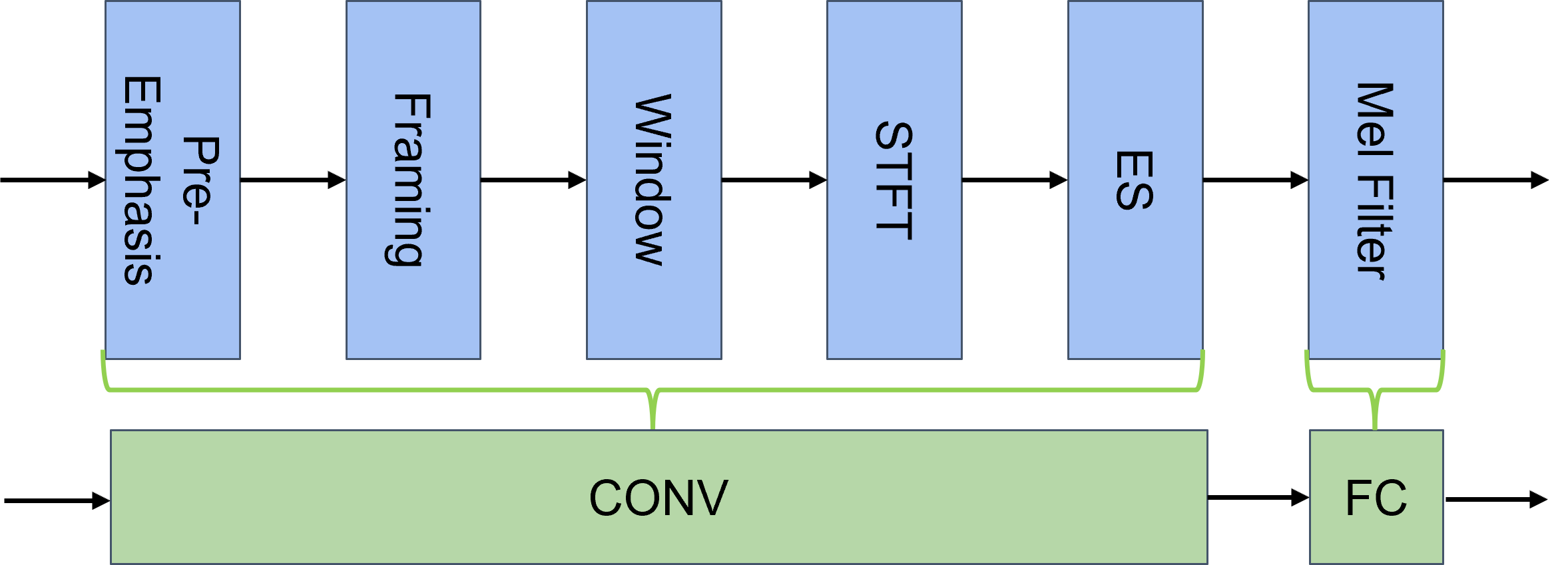}
	\caption{The framework of \textit{time-domain feature extractor}. The processes in the standard mel-fbank feature extractor, including pre-emphasis, framing, window, STFT and ES are replaced by a convolutional layer, and the mel filter is substituted with a fully-connected layer.}
	\label{fig:convfbank}
\end{figure}

In general, the standard mel-fbank feature extractor consists of following modules: pre-emphasis, framing, window, short-time Fourier transform(STFT), energy spectrum(ES) and mel filter.  As is conveyed by Fig.\ref{fig:convfbank}, which is the framework of TDFE, all these parts are combined into a black box operation of a convolutional layer and a fully-connected layer.

As features are extracted via a convolutional layer and a fully-connected layer in TDFE, we denote the input tensor as $\mathbf{X} \in \mathbb{R}^{A \times T \times D}$, where $A$ is the amplititude number, $T$ is the time dimension and $D$ is the dimension of input features.  The convolution operation convolves with a weight matrix $\mathbf{W}_{c} \in \mathbb{R}^{1 \times K \times D \times O}$, where $K$ is the kernel size and $O$ is the dimension of output features. Likewise, a weight matrix $\mathbf{W} \in \mathbb{R}^{A \times T \times O \times D}$ is applied in the fully-connected operation.
Mathematically, the process of TDFE could be formulated as:
\begin{equation}
\mathbf{E} = \mathcal{F}(\mathbf{X} \ast \mathbf{W}_{c})
\end{equation}
\begin{equation}
\mathbf{Y} = \mathcal{F}(\mathbf{W} \times \mathbf{E})
\end{equation}
where $\ast$  denotes the convolution operation and $\times$ denotes the matrix multiply, $\mathcal{F(\cdot)}$ represents the activation function, $\mathbf{Y}$ is the output feature likes the mel-fbank feature.

It is worth noting that the general steps to obtain the mel-fbank feature from a speech signal are inspired by the biology of human perception. However, features extracted by our proposed TDFE are learned from the distribution of real-world data, which are more suitable for subsequent processing of neural networks.


\subsection{Diffluence Loss}
\label{sec:difloss}
To address restrictions for capacity and discrimination of speaker embeddings since the original transformer structure may have frame-level information waste on output features, the \textit{diffluence loss} is designed. As we mentioned before, the \textit{diffluence loss} is indicated the distance between the utterence-level embedding and other frame-level embeddings in each layer. Thus, it could enhance the speaker-related information in the utterence-level embedding. 

The loss function for training consists of following two parts, classification loss $L_{C}$ and \textit{diffluence loss} $\mathfrak{L}_{D}$, as formulated in Eq.(\ref{eq:loss}).

\begin{equation}
	\label{eq:loss}
	L = L_{C} - \mathfrak{L}_{D}
\end{equation}

Therein, the classification loss is a cross-entropy function in the standard classification problem.
Here, we use the AAM-Softmax loss function~\cite{deng2019arcface}, as follows:

\begin{equation}
\label{eq:aams}
\begin{aligned}
L_{C} &= -\frac{1}{N} \sum^{N}_{n=1} \log{\frac{\phi_{n}}{\phi_{n} + \psi_{n}}}\\
\phi_{n} &= e^{\tau \cos(\theta_{l_{n},n}+m)}\\
\psi_{n} &= \sum^{J}_{j=1,j \neq l_{n}} e^{\tau \cos(\theta_{j,n})}
\end{aligned}
\end{equation}
where $\tau$ is a scaling factor for preventing gradients too small during the training process, $m$ is the hyper-parameters. 


For the \textit{diffluence loss} $\mathfrak{L}_{D}$, the Kullback-Leibler divergence, which is widely applied in the field of machine learning and deep learning, is utilized as a representation of distance:

\begin{equation}
\label{eq:diffl}
\mathfrak{L}_{D} = \frac{1}{LT} \sum^{L}_{l=1}\sum^{T}_{i=1} KL(\mathbf{v}_{l0} || \mathbf{v}_{li})
\end{equation}
where the $\mathbf{v}_{li}$ represents the $i$-th output embedding in the $l$-th layer, \textit{KL}$(\cdot)$ denotes the Kullback-Leibler divergence.
Specially, when $l = 1$, the $KL(\mathbf{v}_{l0} || \mathbf{v}_{li})$ is called \textbf{attention distance}. 

Besides, the cosine function could also serve as  $\mathfrak{L}_{D}$, which is an another common measurement of the distance between distributions. The relevant results of comparative experiment can be found in Section \ref{sec:exp}.

\section{EXPERIMENT AND ANALYSIS}
\label{sec:exp}
In this section, we firstly describe some datasets utilized in following experiments, and then detailedly analyze the experimental results accordingly.

\subsection{Dataset}

\textbf{CN-CELEB}: CN-Celeb~\cite{fan2020cn} is a large-scale speaker recognition dataset collected ``in the wild". Specifically, the dataset contains more than 130000 utterances from 1000 Chinese celebrities, and covers 11 different genres in real world, including entertainment, interview, singing, play, movie, vlog, live broadcast, speech, drama, recitation and advertisement. The speech of a particular speaker may be in more than 5 genres. The diversity in genres makes CN-Celeb more representative for the true scenarios in unconstrained conditions, but also more challenging.

\textbf{VoxCeleb2}: VoxCeleb is a large-scale speaker identification dataset consisting of short clips of human speech. The entire dataset involves two parts: VoxCeleb1~\cite{nagrani2017voxceleb} and VoxCeleb2~\cite{chung2018voxceleb2}. VoxCeleb2 dataset consists over 1 million utterances for over 6,000 celebrities, extracted from videos uploaded to YouTube. VoxCeleb1 dataset consists of about 150,000 utterances from 1251 different celebrities. The utterances are collected from YouTube videos in which the celebrities belong to different races and have a wide range of accents. In our work, the proposed DT-SV is trained on the VoxCeleb2 while evaluated on the VoxCeleb1, including VoxCeleb1-E and VoxCeleb1-H test sets.

\subsection{Implementation details}
\label{sec:details}

In order to compare experimental results equitably, all settings in our experiments are consistent with those in baselines\cite{zeinali2019improve}, except for the loss function.
Thus, we utilize similar network structure, data processing, training, and testing strategies in our experiments.

\textbf{Network structure}:  As we mentioned in Section \ref{sec:intro}, two different architectures are proposed in our work.  To be specific, DT-SV-light is ``lighter" on three aspects including the number of layers, dimensions and attention heads than DT-SV, as presented in Table~\ref{tab:structure}. Consequently, the number of parameters of DT-SV-light is much less than DT-SV, which makes DT-SV-light much easier to apply in more scenarios like mobile devices.

\textbf{\begin{table}[!htbp]
		\caption{Model hyper-parameters of DT-SV-light and DT-SV.}
		\label{tab:structure}
		\centering
		\begin{tabular}{l|c|c}
			\hline
			\textbf{Model} & \textbf{DT-SV-light} & \textbf{DT-SV} \\
			\hline
			\hline
			\textbf{Params(M)} & 1.0 & 21.5 \\
			\textbf{Layers} & 4 & 6 \\
			\textbf{Dims} & 128 & 128/256/512 \\
			\textbf{Attention Heads} & 4 & 8 \\
			\hline
		\end{tabular}
\end{table}}

\textbf{Metric}: 
Following the general practice,  equal error rate (EER) and minimum detection cost function (minDCF) are chosen to evaluate the performance.  Some parameters maintain the same as those in \cite{nagrani2017voxceleb}, where the target probability $P_{tar}$ is 0.01, $C_{fa}$ and $C_{fr}$ has the same weight of 1.0.

\textbf{Training}: In our design for ablation experiments, four different models with different layers or dimensions are trained to assess the effects of hyperparameters. All training settings are presented in Table \ref{tab:setting}, one of these models has 4 layers and 128 dimensions, while the other three models all have 6 layers with exponentially increasing dimensions(128, 256 and 512 respectively). For the two models with 128 dimensions, the learning rate decay and the weight decay are both set as 0.97 and $10^{-5}$  respectively while the learning rate are 0.001 and 0.0005. For the other two models with 256 dimensions and 512 dimensions, the learning rate, the learning rate decay and the weight decay are basically the same with 0.001, 0.99 and $5 \times 10^{-5}$ respectively. 
To make a relatively fair comparison, the number of epoches is progressive increase because of different model sizes.  Besides, for all models, the optimizer is Adam \cite{kingma2015adam} and the batch size is set to 150.


\begin{table}[!htbp]
	\centering
	\caption{Training setting of different DT-SV models. ``L'' is layer, ``D'' is dimension, ``LR'' is learning rate, ``LRD'' is learning rate decay, ``WD'' is weight decay.}
	\label{tab:setting}
	\begin{tabular}{cccccc}
		\hline
		\textbf{\#L} & \textbf{D}  & \textbf{LR} & \textbf{LRD} & \textbf{WD} & \textbf{epoch}\\
		\hline
		\hline
		4 & 128 & 0.001 & 0.97 & $10^{-5}$ & 20\\
		6 & 128 & 0.0005 & 0.97 & $10^{-5}$ & 24\\
		6 & 256 & 0.001 & 0.99 & $5 \times 10^{-5}$ & 26\\
		6 & 512 & 0.001 & 0.99 & $5 \times 10^{-5}$ & 30\\
		\hline
	\end{tabular}
\end{table}

\textbf{Different training techniques}: Table~~\ref{tab:TIMIT} shows the progressive improvements of DT-SV while employing different training techniques. To verify the effectiveness of common techniques, \textit{diffluence loss} is not added for the time being.
Starting from DT-SV-light, which is regard as baseline, with 3.19\% EER, the multiple training techniques are applied to improve the performence. The first one is SpecAugment \cite{park2019specaugment} and Data Augmentation that could make the distribution of the input raw wave data more diverse. Thus, it lowers the EER to 2.84\%, or reduces 0.35\% EER compared to the baseline. 
Then, with the help of Relative Position Embedding \cite{dai2019transformer} that could enhance the information of relative position, the EER of the model decreases by another 0.5\%. In general, making a model bigger and deeper is always an effective method for acquiring better representations. As shown in Table~\ref{tab:TIMIT}, the model with higher multi-layer perceptron(MLP) dimensions and more Transformer blocks could reduce the EER greatly. Consequently, all these four training techniques are employed in the following experiments.

\begin{table}[!htbp]
	\caption{Ablation path of our proposed DT-SV-light baseline. 
		All the models are carried out on the same GPU devices for fair comparison. 
		The performance can be boosted from 3.19 to 2.07 (\textcolor{red}{-1.12}) using the proposed techniques.
	}
	\label{tab:TIMIT}
	\centering
	\begin{tabular}{l|c}
		\hline
		\textbf{Training Techniques} & \textbf{EER(\%)} \\
		\hline
		\hline
		Basline(DT-SV-light with AAM-Softmax) & 3.19 \\
		+ SpecAugment \& Data Augmentation & 2.84 (\textcolor{red}{-0.35})\\
		+ Relative Position Embedding & 2.69 (\textcolor{red}{-0.50})\\
		+ Higher MLP dimension ($256 \rightarrow 512$) & 2.13 (\textcolor{red}{-1.07})\\
		+ More Transformer Blocks ($4 \rightarrow 6$) & 2.07 (\textcolor{red}{-1.12})\\
		\hline
	\end{tabular}
\end{table}


\begin{table*}[htbp]
	\caption{Comparison of different models on CN-CELEB.}
	\label{tab:diffe2e}
	\centering
	\begin{tabular}{l|c|c|c|c|c}
		\hline
		\textbf{Front-end} & \textbf{Input} &\textbf{Params(M)}&\textbf{GFLOPs}& \textbf{EER(\%)} & \textbf{minDCF} \\
		\hline
		\hline
		x-vector~\cite{fan2020cn} & MFCC & 4.2 & 2.164 & 14.78 & 16.51 \\
		r-vector~\cite{chen2021self} & Fbank & - & - & 13.43 & N/R \\
		Fast-ResNet34~\cite{chung2020defence} & Raw waveform  & 1.9 & 5.874 &13.12 & 0.611 \\
		\hline
		DT-SV-light & Raw waveform & \textbf{1.0} & \textbf{0.390} & \textbf{12.54} & \textbf{0.562} \\
		\hline
	\end{tabular}
\end{table*}
\begin{table*}[htbp]
	\footnotesize
	\caption{Results for speaker verification on the VoxCeleb1, VoxCeleb1-E and VoxCeleb-H test sets. N/R : Not report results. AP: Angular Prototypical. DLoss represents the \textit{diffluence loss}.}
	\label{tab:vsSOTA}
	\centering
	\begin{tabular}{c|c|c|c|c|c|c|c|c}  
		\hline  
		\multicolumn{1}{c|}{\multirow{2}{*}{Front-end Model}}&\multicolumn{1}{c|}{\multirow{2}{*}{Input}}&\multicolumn{1}{c|}{\multirow{2}{*}{Loss}}&\multicolumn{2}{c|}{VoxCeleb1}&\multicolumn{2}{c}{VoxCeleb1-E}&\multicolumn{2}{|c}{VoxCeleb1-H}\cr\cline{4-9}&&&EER(\%)&MinDCF &EER(\%)&MinDCF &EER(\%)&MinDCF\cr 
		\hline
		\hline
		Thin-ResNet34 \cite{nagrani2020voxceleb} & Magnitude & Softmax & 2.87 & N/ R & 2.95 & N/R & 4.93 & N/R\\
		RawNet2 \cite{jung2020improved} & Raw waveform & Softmax & 2.48 & N/R & 2.57 & N/R & 4.89 & N/R \\
		Wav2spk \cite{lin2020wav2spk} & Raw waveform & AM-softmax & 1.95 & 0.203 & N/R & N/R & N/R & N/R \\
		raw-x-vector \cite{zhu2020raw} & Raw waveform & AM-softmax & N/R & N/R & 2.64 & N/R & 4.34 & N/R \\
		ICspk \cite{peng2021icspk}  & Raw waveform & AP & \textbf{1.92} & \textbf{0.137} & \textbf{1.94} & \textbf{0.141} & \textbf{3.78} & \textbf{0.237} \\
		\hline
		TESA \cite{nj2021s} & MFCC & Softmax & 2.37 & 0.2 & N/R & N/R & N/R & N/R \\
		serialized attention \cite{zhu2021serialized} & MFCC & cross entropy & N/R & N/R & 2.36 & 0.242 & 3.99 & 0.375 \\
		\hline
		DT-SV & Raw waveform & AAM-softmax & \textbf{2.07} & \textbf{0.154} & \textbf{2.11} & \textbf{0.161} & \textbf{4.31} & \textbf{0.245}\\
		DT-SV & Raw waveform & AAM-softmax+DLoss(Cosine) & \textbf{1.95} & \textbf{0.132} & \textbf{1.94} & \textbf{0.138} & \textbf{3.96} & \textbf{0.247}\\
		DT-SV & Raw waveform & AAM-softmax+DLoss(KL) & \textbf{1.92} & \textbf{0.130} & \textbf{1.91} & \textbf{0.136} & \textbf{3.72} & \textbf{0.225}\\
		\hline
	\end{tabular}
\end{table*}

\subsection{Results and Analysis}
\label{sec:anal}
\begin{figure}[ht]
	\begin{minipage}[t]{\linewidth}
		\centering
		\includegraphics[width=0.8\linewidth]{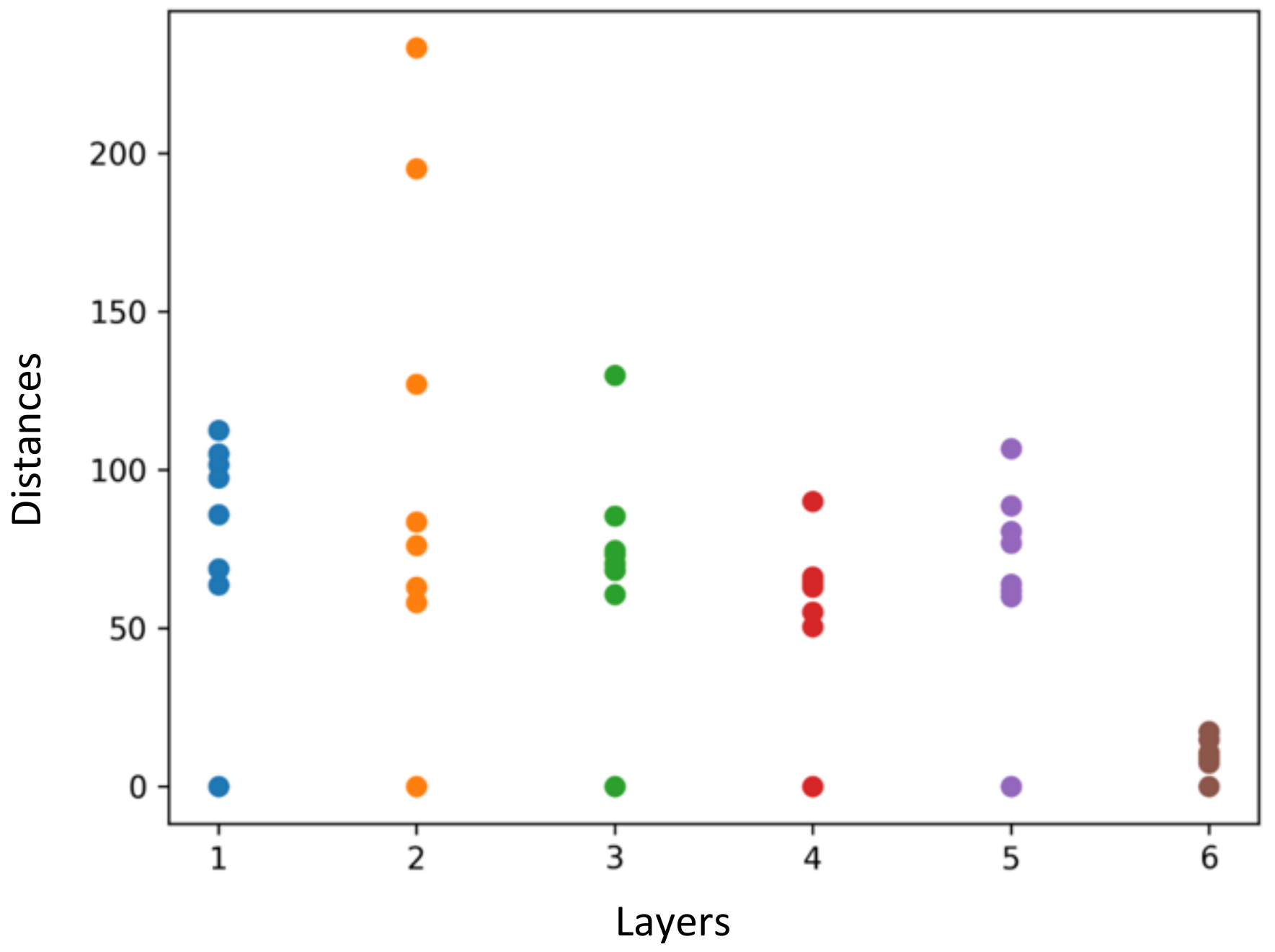}
		\centerline{(a) Attention distance when the layer of Transformer is 6.}\medskip
	\end{minipage}
	\hfill
	\begin{minipage}[t]{\linewidth}
		\centering
		\includegraphics[width=0.8\linewidth]{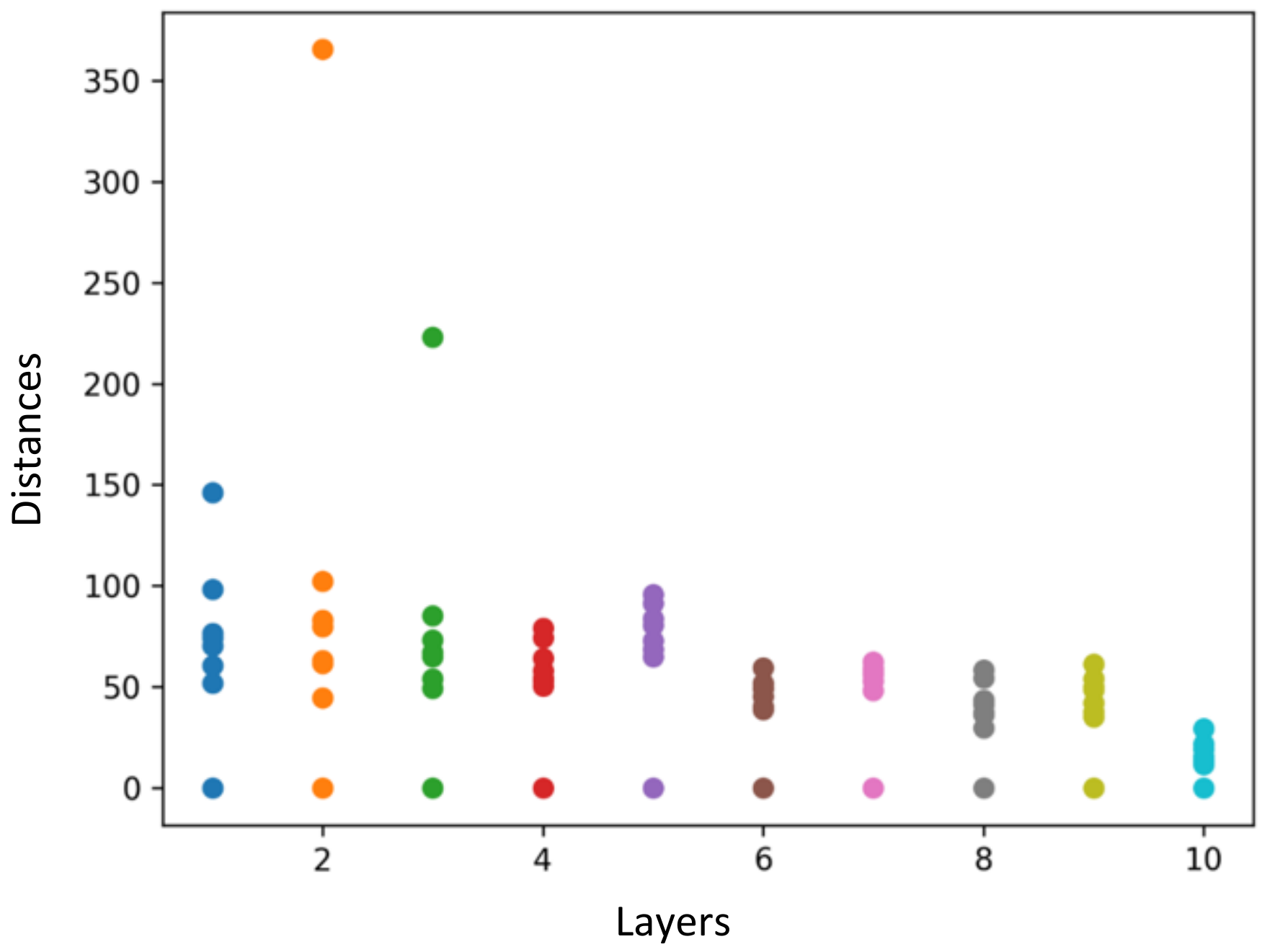}
		\centerline{(b) Attention distance when the layer of Transformer is 10. }\medskip
	\end{minipage}
	\caption{The attention distance in each layer.}
	\label{fig:attn}
\end{figure}

\textbf{Attention distance}: 
Here the impact of layers on performance of DT-SV is explored. The Fig.~\ref{fig:attn} displays the distance between attention heads at different layers. Therein, dots with the same color represent the distance between the first attention head and the other attention heads in the specific layer. The  distance of attention head is calculated by averaging distances of the first frame embedding and the other frame embeddings, as Eqn. (\ref{eq:diffl}) in Sec.~\ref{sec:difloss} when $l=1$. 

As illustrated in Fig.~\ref{fig:attn}(a), when the layer of Transformer is 6, the attention distance of the first five layers is relatively scattered. Following the second layer that is the most scattered, the degree of scattering from the third layer to the fifth layer is similar to the first layer. The attention distances are the most clustered in the sixth layer. Moreover, as shown in Fig.~\ref{fig:attn}(b) in which the number of layers of Transformer is 10, the overall trend is basically similar to Fig.~\ref{fig:attn}(a). Specifically, the attention distances in the first five layers are relatively scattered, and the distance is gradually gathering from the sixth layer.

From the whole trend of attention distance in Fig.~\ref{fig:attn}, we notice that each layer is broadly more clustered than its former layer from the second layer, which indicates that \textit{diffluence loss} could allow DT-SV to integrate information across the entire utterance to the lower layers. And the distance is stable from the sixth layer, that is to say, the differences among the frame-level embeddings in these layers is slight. The results inspire us that the number of layers of the Transformer in the DT-SV could be set as 4 to hold fewer parameters and achieve decent performance. Alternatively, such number could also be set as 6 to obtain a more powerful model.

\textbf{Comprehensive comparison:}
The metrics of different models on the CN-CELEB are listed in Table~\ref{tab:diffe2e}. Notice that we select the DT-SV-light to compared with the other three models, wherein the inputs of x-vector and r-vector are the MFCC or Fbank features extracting from the waveform by the standard preprocessing in Fig.\ref{fig:convfbank} while the input of the Fast-ResNet34 is the raw waveform.
As presented in Table~\ref{tab:diffe2e}, the DT-SV-light achieves the best performance with 12.54\% EER, which proves that the TDFE could acquire much better features than the standard preprocessing. Moreover, the fact that the DT-SV-light could achieve decent performance with fewer parameters suggests its enormous potential in industrial applications.

To futher demonstrate the effectiveness of our proposed model, we conduct more comprehensive comparison experiments, as the Table~\ref{tab:vsSOTA} shows, in which the other results are cited from the literatures. Therein, the input of these models are raw waveforms except Thin-ResNet34, TESA and serialized attention model, though TESA and serialized attention are also based on Transformer. 

Obviously, the proposed DT-SV could achieve lower EER than most comparative models on all datasets even without the \textit{diffluence loss}. Furthermore, the better performance of the DT-SV after combining with the \textit{diffluence loss}, especially that calculated by KL divergence, indicates that the \textit{diffluence loss} could enhance the related-speaker information effectively. Besides, in comparison to TESA and serialized attention that based on Transformer with well-designed inner structures, the proposed DT-SV based on original Transformer with some minor modifications, such as \textit{diffluence loss}, is quite simple yet extremely effective. More significantly, such loss could be easily incorporated into any existing Transformer-based achitectures.

\section{Conclusion}
In this work, we propose an improved Transformer-based time-domain approach for speaker verification named DT-SV, which leverages \textit{time-domain feature extractor}  and \textit{diffluence loss} to capture speaker characteristics in an utterance having greater capacity and discrimination. Specifically, the \textit{time-domain feature extractor} could extract features from the raw speech input more precisely and efficiently than the standard mel-fbank extractor. Moreover, a novel loss function named \textit{diffluence loss} could enhance speaker-related information in the utterance-level embedding while weaken such information in other frame-level embeddings. As our experiments demonstrated, the proposed DT-SV could achieve competitive results on the datasets employed. Besides, our proposed architecture explores the Transformer mechanism for the speaker verification task and hopefully gives an inspiration for other related tasks. Last but not least, the \textit{diffluence loss} could be flexibly embedded into other analogous structures.


\section*{Acknowledgment}
This paper is supported by the Key Research and Development Program of Guangdong Province under grant No.2021B0101400003. Corresponding author is Jianzong Wang from Ping An Technology (Shenzhen) Co., Ltd (jzwang@188.com).
\bibliographystyle{IEEEtran}
\bibliography{spkformer}

\end{document}